# New $Tl_2LaBr_5$: $Ce^{3+}$ Crystal Scintillator for γ-Rays Detection

H. J. Kim[a*] Gul Rooh[b], Arshad Khan[a], Sunghwan Kim[c]

[a]*Department of Physics, Kyungpook National University, Daegu 41566, Korea*
[b]*Department of Physics, Abdul Wali Khan University, Mardan, 23200, Pakistan*
[c]*Department of Radiological Science, Cheongju University, Cheongju 41566, Korea*

**Abstract**

In this study we present our preliminary report on the scintillation properties of new Ce-doped $Tl_2LaBr_5$ single crystal. Two zones vertical Bridgman technique is used for the growth of this compound. Pure and Ce-doped samples showed maximum emission peaks at 435 nm and 415 nm, respectively. Best light yield of 43,000±4300 ph/MeV with 6.3% (FWHM) energy resolution is obtained for 5% Ce-doped sample under γ-ray excitation. Single exponential decay time constant of 25 ns is observed for 5% Ce doped sample. Effective Z-number is found to be 67, therefore efficient detection of X- and γ-ray will be possible. Preliminary results revealed that this compound will be an ideal candidate for the medical imaging techniques. Further investigations are under way for the determination of optimized conditions of this compound.

*Corresponding author: Tel.:+82-53-950-5323; fax: +82-53-956-1739. E-mail address: hongjoo@knu.ac.kr (H.J. Kim).

**1.0 Introduction**

For the efficient detection of ionizing radiations, radiation detector based on inorganic scintillation compounds are used in many applications [1]. Different scintillators are discovered and employed in various application since the discovery of NaI: Tl [2, 3]. Most of the research and development for the discovery of new scintillation compounds is devoted to achieve a superior scintillation compound which can be used in all applications. Unfortunately, such an ideal scintillator could not be discovered so far. Usually a scintillation compound should possess high light output, good energy resolution, fast scintillation response, high density and high effective Z-number [4]. Inorganic halide scintillators are considered to be the best among the discovered scintillators. Specially, Ce-doped inorganic halide scintillators are extensively studied and their scintillation properties are found excellent among the discovered scintillators [5, 6].

Recently, we discovered new class of inorganic halide scintillators which contained Thallium (Tl) ion and activated with $Ce^{3+}$-ions. Due to high density and high Z-number of Tl ($\rho$ = 11.8 g/cm$^3$, Z = 81) ion, these scintillators showed excellent scintillation performance [7-11]. These scintillators showed excellent performance and could be used in different applications such as medical imaging techniques, high energy and nuclear physics research, homeland security and basic research in condense matter physics.

In this study, we report on new $Tl_2LaBr_5$: $xCe^{3+}$ where x = 0 and 5 mole % (TLB). For the growth, two zones vertical Bridgman technique is used. Luminescence and scintillation properties are measured under X-ray and γ-ray source at room temperature. Luminescence properties include X-ray induced emission spectra. Scintillation properties of TLB crystal include energy resolution, light output and decay time.

## 2.0 Experimental section

### 2.1 Crystal growth

Pure and 5%Ce doped single crystals of TLB has been grown by two zone vertical Bridgman technique. Stoichiometric amounts of TlBr (99.999%, Alfa-Aesar), $LaBr_3$ (99.999%, Sigma-Aldrich) and $CeBr_3$ (99.999%, Sigma-Aldrich) powders were weighing and loaded in quartz ampoule inside argon purged glovebox. Powder loaded ampoules were sealed under high vacuum and were grown by two zone vertical Bridgman technique. Grown crystals were transparent, homogenous and cracks free. The grown ingots were cut into slices of Ø 8 x 2 mm$^3$ dimensions with diamond cutting wire followed by polishing with polishing cloth. Figure 1 shows the as grown and polished sample of TLB single crystal. Grown samples were kept in mineral oil to avoid the surface degradation. Density and effective Z-number of TLB is found to be 5.9 g/cm$^3$ and 67, respectively.

### 2.2 Experimental setup

X-ray induced luminescence spectra of TLB: $Ce^{3+}$ crystals was measured with an X-ray tube having W anode from a DRGEM. Co at room temperature. Power setting of the X-ray generator was set as 100 kV and 1 mA. A QE65000 fiber optic spectrometer made by Ocean Optics was used to measure the X-ray induced emission spectra of the crystals. For the evaluation of pulse height spectra, TLB: $Ce^{3+}$ crystals were wrapped in several layers of Teflon tape with one face left uncovered. The uncovered face of the crystal was directly coupled with the entrance window of the photomultiplier tube (PMT) (R6233, Hamamatsu) using index matching optical grease. After irradiated with 662 keV γ-rays from a $^{137}Cs$ source, the signal of the detector were shaped with a Tennelec TC 245 spectroscopy amplifier and then fed into a 25-MHz flash analog-to-digital converter (FADC) [16=12]. A software threshold was set to trigger an event by using a self-trigger algorithm on the field programmable gate array (FPGA) chip of the FADC board. The FADC output was recorded into a personal computer by using a USB2 connection, and the recorded data were analyzed with a C++ data analysis program [13].

For the light output measurement of TLC crystals, a calibrated LYSO: $Ce^{3+}$ (light yield = 33,000 ph/MeV) was used. Similar experimental set up of pulse height measurement was used for the light out measurement. TLB and LYSO crystals were wrapped in Teflon tap and mounted on the entrance window of the (PMT) (R6233, Hamamatsu). Both crystals were excited with $^{137}Cs$ γ-ray source by keeping similar conditions of PMT bias and amplifier gain. Light output of TLB crystal was obtained by comparing the channels numbers of the recorded pulse height spectra of both crystals [14].

For the decay time measurements of the TLC crystals, PMT (R6233, Hamamatsu) signals generated due to 662 keV γ-ray excitation in the crystals were digitized by 400 MHz FADC. FADC module (homemade) was designed to sample the pulse every 2.5 ns for duration up to 64 μs [15]. A trigger was formed in the FPGA chip on the FADC board. For low energy events, more than four photoelectrons in 2 μs were needed for the event trigger. An additional trigger was generated if the width of the pulse was longer than 200 ns for high energy events where many single photon signals were merged into a big pulse. The FADC located in a VME crate was read out by a Linux-operating PC through the VME-USB2 interface with a maximum data transfer rate of 10 Mbytes/s. Decay time of TLB was found from the recorded pulse shape information [15].

**3.0 RESULTS AND DISCUSSION**

**3.1 X-ray induced luminescence**

TLB pure and 5%Ce doped crystals X-ray induced luminescence spectra is shown in Fig. 2. For pure sample, emission band between 350 nm and 525 nm peaking at 435 nm is observed. 5%Ce doped sample showed typical Ce$^{3+}$ emission band between 350 nm and 460 nm peaking at 375 nm and 415 nm. In the 5%Ce sample two overlapped peaks at 375 nm and 415 nm is assigned to the 5d → $^2F_{5/2}$ and 5d → $^2F_{7/2}$ transition of Ce$^{3+}$ ion, respectively. Various peaks observed above 525 nm in the pure sample is assigned to an unknown impurity in TlBr and LaBr$_3$ powders. No such peaks are observed in the 5%Ce crystal. Origin of the luminescence in the pure crystal is unknown however, it might be due to the localized Tl$^+$ ion or some intrinsic defect in TLB crystal [16, 17].

## 3.2 Pulse height spectra and light yield

$^{137}$Cs-γ-ray pulse height spectra of TLB: Ce$^{3+}$ crystals is illustrated in Fig. 3(a)-(b). For comparison, the pulse height spectrum of LYSO: Ce$^{3+}$ crystal is also included in Fig. 3(a)-(b). In the Fig 3, the observed photopeak positions (channels number) is proportional to the light output. Comparing the photopeak channel numbers of the pure and 5%Ce doped TLB with LYSO: Ce$^{3+}$ crystal, the corresponding light outputs are found to be 16,000±1600 ph/MeV and 43,000±4300 ph/MeV, respectively. The energy resolutions are obtained by using Gaussian fits to the photopeaks of the pure and 5%Ce doped samples. The energy resolutions of the pure and 5%Ce doped samples are obtained to be 8.1% (FWHM) and 6.3% (FWHM), respectively. In Fig. 3(a)-(b), the pulse height spectra of pure and 5%Ce doped samples showed pronounced low energy peaks between 1100 to 1200 and 1520 to 1845 channel numbers, respectively due to escape peaks of K$_β$ and K$_α$ X-rays from Tl atoms of the TLB crystals [7]. From the obtained energy resolution and light output, it is evident that addition of Ce$^{3+}$ ion in the host improved the scintillation properties of TLB crystal. Therefore, it is expected that with optimized condition of Ce-concentration, further improvement is possible in this host lattice.

## 3.3 Scintillation decay time

Figure 6 illustrates the decay time spectra of TLB: Ce$^{3+}$ crystals under γ-ray excitation at room temperature. After fitting the decay curves with single and two exponential functions, the pure sample showed two exponential decay time constants of 84 ns (66%) and 172 ns (34%). 5%Ce doped exhibits single exponential decay time constant of 25 ns. Inset of Fig.4 shows the decay time spectrum of the TLB: pure crystal along with the fit to the data and the fast and slow exponential decay constants. The obtained 25 ns decay constant is assigned to the life time of 5d excited state of Ce$^{3+}$ ion. Moreover, the fast scintillation decay time constant revealed that direct electron-hole at luminescence center is possible in TLB: Ce$^{3+}$ single crystal [18]. The effect of Ce$^{3+}$-concentration on the decay time constants of TLB implies that the excitation energy is efficiently transferred from the host lattice to $Ce^{3+}$ ion. Comparing with commercially available scintillators, TLB: Ce3+ could compete in many applications. In addition, this compound has the potential to be used in Positron Emission Tomography (PET) technique.

**Conclusions**

In this study we grew new scintillator TLB: $Ce^{3+}$ single crystals by using two zones vertical Bridgman method. This material exhibits high effective Z-number (67) and density, therefore it has high X- and γ-rays detection efficiency. Pure sample showed luminescence under X-ray excitation between 350 nm and 525 nm. This emission is attributed to the localized Tl+ ion or some intrinsic defect. Ce-doped sample showed typical 5d → 4f emission of $Ce^{3+}$ ion at room temperature. Under $^{137}$Cs-γ-ray excitation, pure and Ce-doped samples showed energy resolutions of 8.1% (FWHM) and 6.3% (FWHM), respectively. Light outputs of the pure and Ce-doped samples are found to be 16,000±1600 ph/MeV and 43,000±4300 ph/MeV, respectively. Single fast decay time constant of 25 ns is obtained for 5%Ce-doped crystal. Our preliminary results revealed that this scintillation compound will be an ideal candidate for the PET application. Further investigations with different Ce-concentrations are under way and we found improvements in the scintillation properties of TLB.

**Acknowledgment**

These investigations have been supported by the National Research Foundation of Korea (NRF) funded by the Ministry of Science and Technology, Korea (MEST) (No. 2015R1A2A1A13001843).**References**

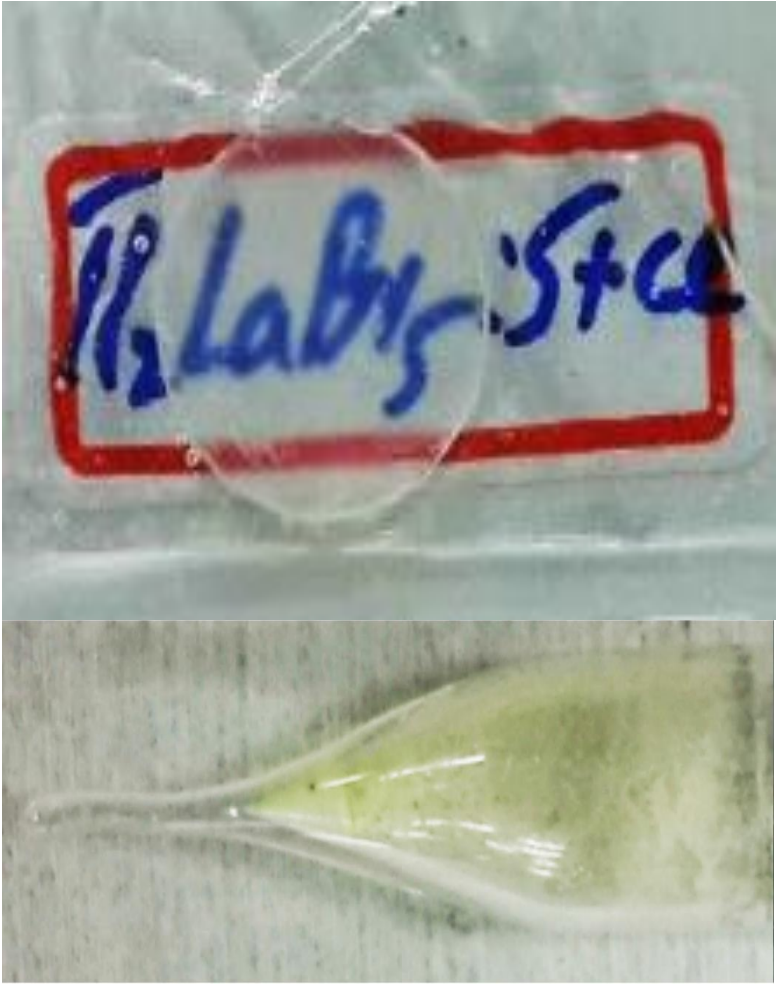

Fig 1. Photograph of the as grown and polished Ø 8 x 2 mm$^3$ TLB single crystal.

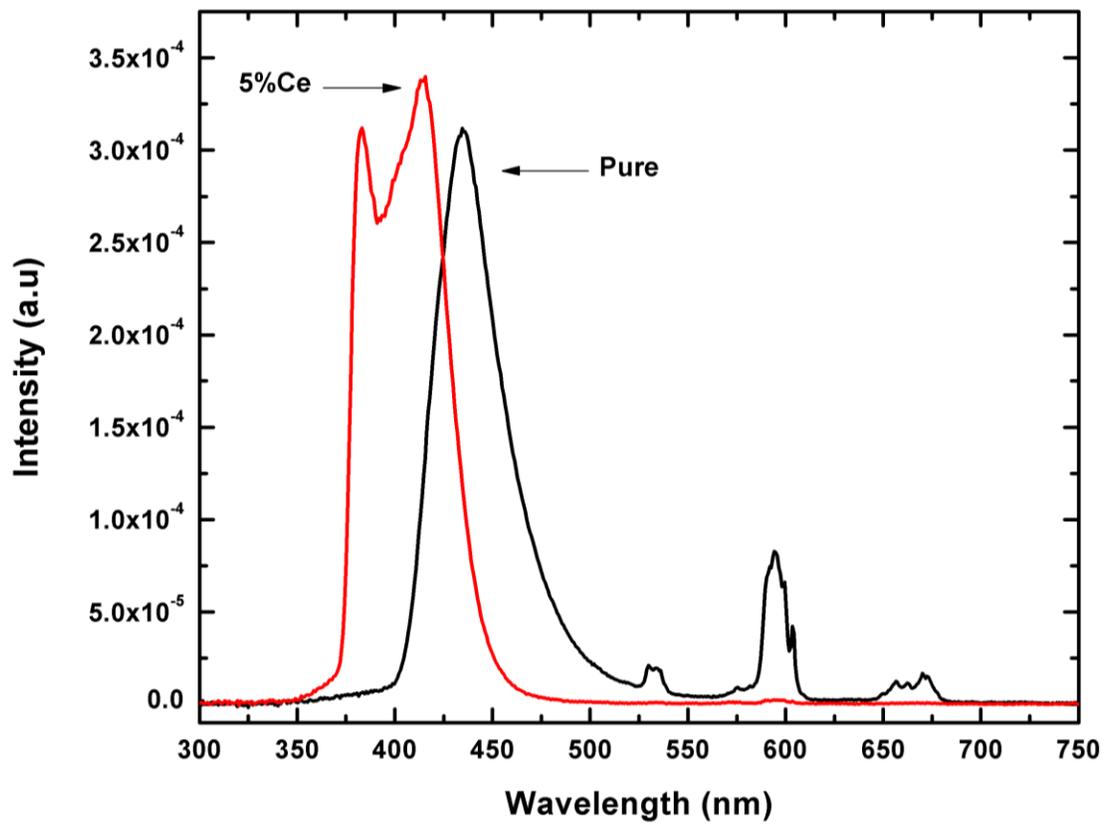

Fig 2. X-rays induced luminescence spectra of pure and 5%Ce doped TLB single crystals at room temperature.

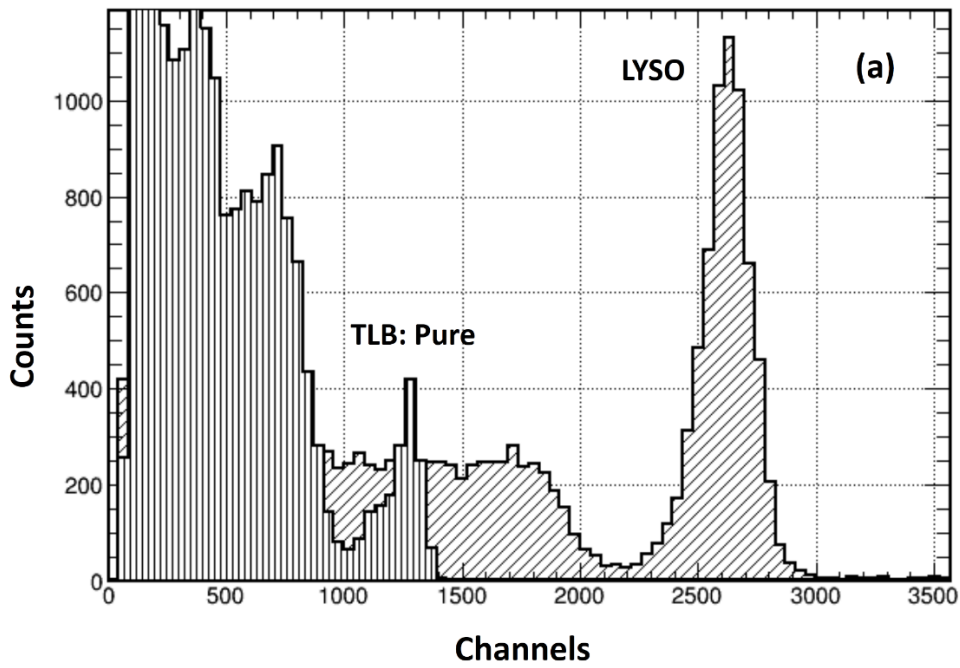

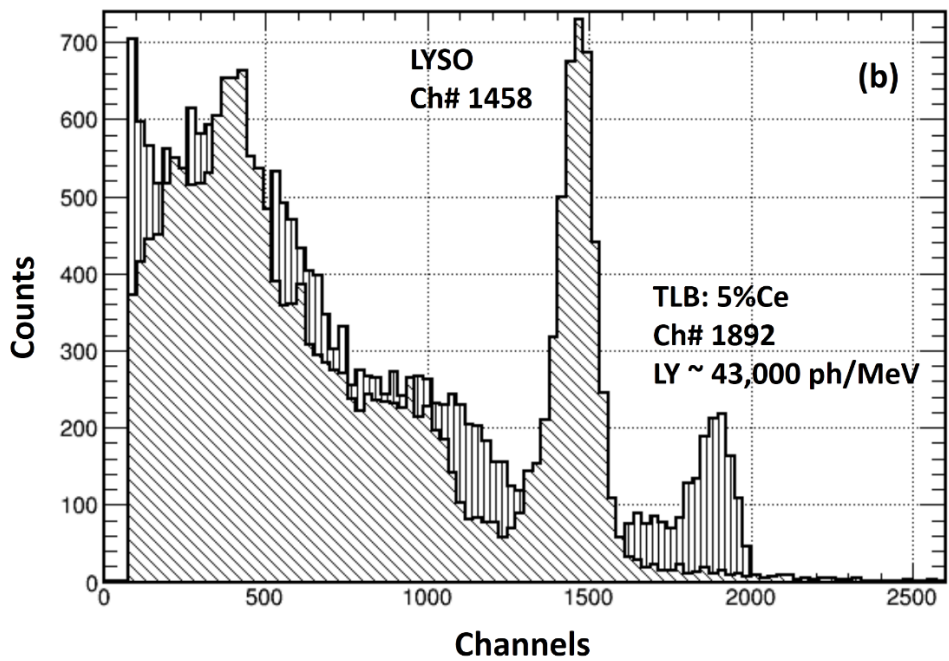

Fig 3. Pulse height spectra of LYSO: $Ce^{3+}$ and TLB (a) pure (b) 5%Ce-doped crystals excited with $^{137}$Cs-γ-rays. The photopeak channel numbers are proportional to the light yield.

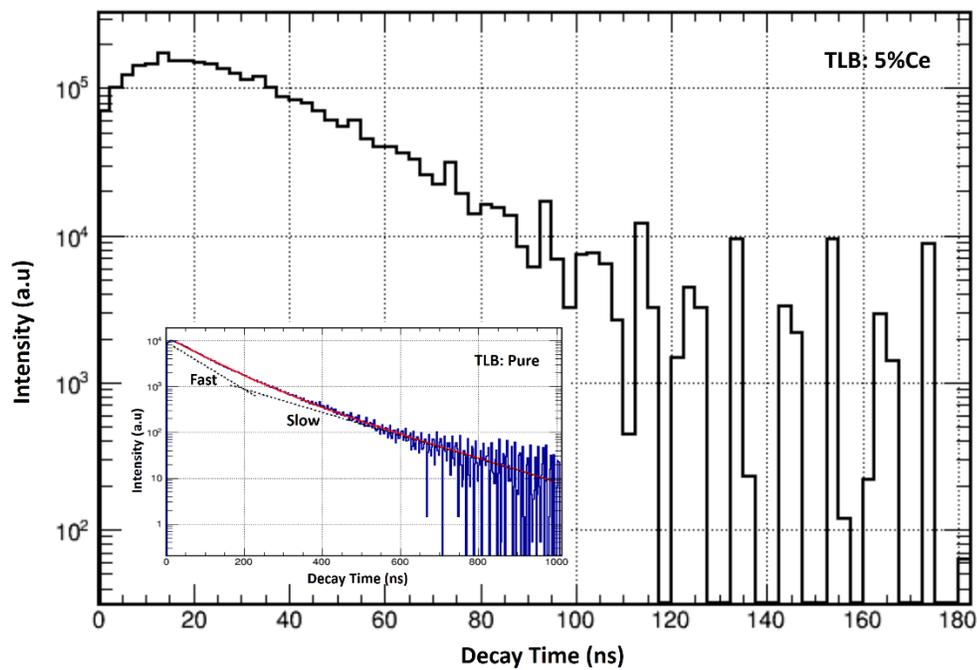

Fig 4. Decay time spectra of TLB: 5%Ce$^{3+}$ at room temperature under $^{137}$Cs- γ-ray excitation. Inset shows TLB: Pure decay time spectrum. Red line shows best fit to the data, fast and slow components is also shown.